\documentstyle [12pt,aasms4]{article}
\input epsf

\journalid{}{}
\articleid{}{}

\begin{document}
\def\lax    {\ifmmode{_<\atop^{\sim}}\else{${_<\atop^{\sim}}$}\fi}
\def\gax    {\ifmmode{_>\atop^{\sim}}\else{${_>\atop^{\sim}}$}\fi}
\def\gtorder{\mathrel{\raise.3ex\hbox{$>$}\mkern-14mu
             \lower0.6ex\hbox{$\sim$}}}
\def\ltorder{\mathrel{\raise.3ex\hbox{$<$}\mkern-14mu
             \lower0.6ex\hbox{$\sim$}}}

\title{The High-Energy Spectra of Accreting Black Holes: Observational
Evidence for Bulk-Motion Infall}

\author{ Chris Shrader\altaffilmark{1,2}
and Lev Titarchuk\altaffilmark{1,3} }

\altaffiltext{1}{Laboratory for High--Energy Astrophysics,
NASA Goddard Space Flight Center, Greenbelt, MD 20771, USA;
shrader@grossc.gsfc.nasa.gov,titarchuk@lheavx.gsfc.nasa.gov}
\altaffiltext{2}{Universities Space Research Association,Lanham MD}
\altaffiltext{3}{George Mason University/Institute for
Computational Sciences and Informatics, Fairfax VA}

\rm

\vspace{0.1in}

\begin{abstract}

 We discuss the emergent spectra from accreting black holes,
considering in particular the case where the accretion is characterized
by relativistic bulk motion. We suggest that such accretion is likely to
occur in a wide variety of black hole environments, where the strong
gravitational field is expected to dominate the pressure forces, and
that this likely to lead to a characteristic high-energy spectroscopic
signature; an extended power-law tail.  It is in
the high (soft) state that matter impinging upon
the event horizon can be viewed directly,
and the intrinsic power-law is seen.
Certain types of Active Galactic Nuclei (AGN) may represent the
extragalactic analog of the high-soft state accretion, which would
further support our ideas, demonstrating the stability of the
($\alpha\sim1.8$) power-law.
This stability
is due to the asymptotic independence of the spectral index on the mass
accretion rate and its weak dependence on plasma temperatures. We
have computed the expected spectral energy distribution for an
accreting black hole binary in terms of our three model parameters: the
disk color temperature, a geometric factor
related to the illumination of
the black hole site by the disk and a spectral index related to the
efficiency of the bulk motion upscattering.
We emphasize that this is a fully self-consistent approach, and
is not to be confused with the more common phenomenological
methods employing additive power law and black-body or
multi-color disk. A test of the model is
presented using observational data from the Compton Gamma Ray
Observatory and the Rossi X-Ray Timing Explorer, covering $\simeq
2-200$ keV for two recent galactic black hole X--ray nova outbursts.
The resulting model fits are encouraging and, along with some
observational trends cited from the literature, they support our
bulk-motion hypothesis.

\end{abstract}


\section{INTRODUCTION}

Do black holes interact with an accretion flow in such a way that a
unique observational signature can be identified -- that is one  which
is entirely distinct from those associated with other compact objects,
based solely on the radiation observed at infinity? This is a crucial
question
confronting both theoretical and observational astrophysicists today;
for recent reviews of the astrophysics of black holes [see e.g.,
\cite{liang97}, \cite{zhang97a}]. Certainly a large body of
evidence has been accumulated which supports the {\it existence} of
black holes, the most convincing arguments being those invoking
dynamical mass determinations [e.g. \cite{orosz97}].  
Other arguments have recently
been advanced suggesting that X-ray nova flux histories demonstrate the
existence of  black-hole horizons [\cite{narayan96}], and in AGN,
asymmetrical line features have identified and interpreted as
originating in massive black hole environments [\cite{tanaka95},
\cite{fabian95}].
Also,  quasi-periodic oscillations (QPOs) have been attributed to
dynamical time scales associated with the innermost stable orbits in
black hole binaries [\cite{mrg97}].

A distinct feature of black hole spacetime geometries, as opposed to
those associated with  other compact objects, is the presence of the
event horizon. Near the horizon the strong gravitational field is
expected  to dominate the pressure forces and thus drive the accreting
material into free fall. In contrast, for other compact objects the
pressure  forces are dominant near the surface and the free fall state
is absent. Recently, Titarchuk and Zannias (1998) (hereafter TZ98)
have developed the relativistic radiative  transfer theory demonstrating
that  high-energy photons are produced by upscattering  from the
converging inflow within a few Schwarzschild radii. Only some  fraction of the
radiation emitted by the accretion disk illuminates the converging
inflow site. It can be such a situation that the radiation density 
(or pressure)  determined
by the injected energy of those soft disk photons and by the weak
amplification they experience [\cite{tmk97}; hereafter TMK97] is much 
smaller than the Eddington value.
{\it We argue that then this difference is crucial and it results in a unique
observational signature for accreting black holes}.
\par
\noindent
As explained above, this signature originates from upscattering of low
energy photons by fast moving electrons  with velocities, $v$,
approaching
the speed of light, $c$. A soft photon of energy $E$,
 in the process of multiple scattering off  the electrons, gets
substantially blue-shifted to energy
\begin{equation}
E^{\prime}=E{{1-(v/c)\cos\theta}\over{1-(v/c)\cos\theta^{\prime}}}
\end{equation}
\noindent
due to  Doppler effect provided at least one photon is scattered in
the direction of electron motion (i.e. when
$\cos\theta^{\prime}\approx 1$).
For example, in the first scattering event we assume 
the direction of
incident photon,  $\theta_1$, is nearly normal to the electron velocity,
 and the direction of the scattered photon is nearly aligned with the
electron velocity. In the process the its outward propagation through the
converging-inflow medium, the angle between the photon and electron
velocity increases. Thus, in the second event the cosine angle, $\cos\theta_2$,
tends to approach zero. The angle of outgoing photon, $\theta_2^\prime$,
has to be large enough,  in order for the Doppler boosted photon
to reach an observer.

Any system having a disk structure around  a compact object
is expected to have a  source of low-energy photons.
{\it The boosted photon component is seen
as the extended power-law at energies much
higher than the characteristic energy of the soft photons.
And it is entirely independent of the initial
spectral and spatial distributions of the low-energy photons.}
The spectral index of the boosted photon  distribution is determined
only by the mass accretion rate and the plasma temperature of
the bulk flow.
{\it The presence of this high-energy power-law  component is a
generic feature of the model.}
A key ingredient in
support of our claim comes from the exact relativistic transfer
calculations describing the Compton scattering of the low-energy
radiation field of the Maxwellian distribution of fast moving
electrons (TZ98). It was proven mathematically that the
power law is always present as a part of the black hole spectrum over 
a wide energy range,
extending up to 500 keV. A turnover in the spectrum at about this
energy, i.e. at E$\ltorder m_ec^2$, is a prediction of our model.
Other extended power-law components, which may be related to the
relativistic electron motion, e.g. in a jet, are not uniquely
constrained to this energy band because they are not tied with the
electron rest mass $m_ec ^2$.
The observations with CGRO/OSSE
could in principle confirm or refute this prediction.
In practice, the data  thus far obtained are signal-to-noise limited and
cannot address this issue in a definitive manner.
\par
\noindent
In this letter we extract the most important points regarding the
radiative transfer and  present observational evidence which 
supports the model.

\section { Bulk-Motion Spectral Models}
It has been shown elsewhere (TMK97, TZ98, \cite{ct95}; hereafter CT95)
that two
effects, the bulk motion upscattering and the Compton (recoil)
downscattering (herein BMC), compete forming the hard tail of the 
spectrum as an
extended power law. The soft part of the spectrum comes from the disk
photons seen directly and a subset of those photons which escape from
the BMC atmosphere after undergoing a few scatterings but
without any significant  energy change. It has also been shown that
without taking  into account special and general  relativistic effects,
one is able to reproduce the main features of the full relativistic
formalism: the overall spectral energy
distribution and the dependence of the high-energy power on
mass-accretion rate (TMK97; TZ98); also, refer to  recent
calculations by Laurent \& Titarchuk (1998) (hereafter LT98).  In the
relativistic treatment, the Compton downscattering becomes less
efficient at high energies, due to Klein-Nishina effects.  Also, there
is a possibility that  the electron distribution in the converging
inflow can deviate from a Maxwellian,  flattening at high velocities,
since there is insufficient time for it to thermalize.
The hard photon power-law thus extends to higher energies.
At the same time however, the spectrum is steepened as a  result of
gravitational redshift effects.

We have assumed that there is an external illumination of
the converging flow by the low-energy black body radiation of an
accretion disk having a characteristic temperature $T_{c}$. Furthermore we
have assumed that this illuminating radiation impinges on
the BMC atmosphere with a certain geometry, which we
have paramaterized in terms of a ''fraction" $ f$.
This fraction is really the first expansion
coefficient of the spatial source photon distribution
over the set of the eigenfunctions of
the BMC formulation (TMK97, Eq. 30).
As we mentioned above (\S 1) the spectral index  is independent of the 
illumination fraction, $f$. It is clearly demonstrated in TMK97 
(Figs 4).

We remind the reader that all reasonable theoretical
spectra must exhibit a smooth transition from blackbody-like spectrum
to a pure power-law, typically, at energies in the 5--12  keV range for
stellar black holes.

In the soft state when the accretion rate is higher, the soft photons
from the the Keplerian disk cool the hot region (Compton cloud)
due to thermal Comptonization and free-free emission (CT95).
The cooler converging inflow, as it rushes towards the black hole, scatter the
soft-photons within the a radius, $r\sim \dot m r_s$ -- some of the
photons  then undergo outward radiative diffusion.
Here $\dot m=\dot M/\dot M_E$, $r_s$ is Scwarzschild radius, $\dot
M$ is the net accretion rate (including accretion from the disk
plus any halo or other non-keplerian component), $\dot M_E \equiv
L_E/c^2=4\pi GMm_p/ \sigma_Tc~$ is the Eddington accretion rate,
$M$ is the mass of the central object, $m_p$ is the proton mass and
$G$ is the gravitational constant. It transfers its momentum to the
soft-photons  to produce the power-law component
extending to energies comparable to the kinetic energy of electrons in
the converging inflow, i.e. of order $m_ec^2$. 
On the other hand {\it in the hard
state, the hot emission cloud covering the BMC zone prevents us 
from seeing the photons that are upscattered to
subrelativistic energies within a few Scwarzschild radii}.

The luminosity of the upscatterd component, the hard power law, has 
to be very small compared to the Eddington luminosity 
in order for the BMC model to be valid. 
The relative normalization of the soft component to the
hard power-law is less important provided the inferred luminosity of 
the hard power-law remains consistent with the assumption 
of negligible radiation pressure near the black-hole horizon.

The BMC spectral model
can be described as the sum
of a thermal (disk) component and the convolution of some fraction of
this component $g(E_0)$ with the upscattering Green's function
$I(E,E_0)$ (TMK97, Eq. 30). The Green's function has the form of a broken
power-law with spectral
indices $\alpha$ and $\alpha+\zeta$ for high  $E\geq E_0$ and low
$E\leq E_0$ energy parts respectively,
\begin{equation} F_{\nu} (E)=\int_0^{\infty}I(E,E_0)g(E_0)dE_0.
\end{equation}
\par
\noindent
The above convolution
is insensitive to the value of
the Green's function spectral index $\alpha+\zeta$, which is always
much greater than one. TZ98 presented rigorous proof that the hard
power-law tail is a signature of a Schwarzschild black hole.
Furthermore, the same statement is valid for the case of a  rotating
(Kerr) black hole, although a higher mass accretion rate is required to
provide the same efficiency for the soft photon upscattering.

We note that the processes of absorption 
and emission (as free-free or synchrotron radiation) can be neglected
provided the plasma temperature of the bulk flow is of order 1 keV 
or greater 
for characteristic number densities 
of order $10^{18}$ cm$^{-3}$ and for magnetic field strengths
in the proximity of the black hole of order $10^4-10^5$ 
gauss or less (CT95).  

\section {Application to  Recent High Energy Observations}
As a test of the model we have collected data resulting from
high-energy observations covering recent activity periods
in two galactic X--ray novae: GRO J1655--40 [\cite{zhang97b}]
and GRS 1915+105 [\cite{chaty96}]. X--ray novae comprise perhaps the
best test case of the methodology described here, since they are in
low-mass binary systems -- avoiding  the added complications which
may arise from the OB star winds in high-mass binary BHCs such as
Cygnus X--1 -- and because they become exceptionally bright in
outburst exhibiting  frequent and pronounced high-energy spectral state
transitions [e.g. \cite{csl97}, \cite{ebisawa94} \cite{esin97}].
Furthermore, as a group they comprise the most convincing galactic
black--hole candidates.

We constructed composite high-energy spectra for GRO J1655--40
during an outburst in the spring of 1996 (\cite{hynes98}), covering the
~2-200 keV spectral region and fit these data by the BMC
model. This was accomplished using summed, standard mode (128
channel) data from the RXTE/PCA and the 16-channel BATSE/LAD
earth-occultation data bracketting the pointed observations.
In addition, there was substantial outburst
activity in GRS 1915+105 during the latter part of 1996 [e.g.
\cite{bandy98}]. We utilized some of the available data from the
same instruments for this event as well.

The BMC model described in section 2 was imported into the
''XSPEC" software package which was used to perform all of the model
fitting described here. Our resulting fits are shown in Figure 1.

For GRO J1655-40 we obtained a blackbody color temperature  of
$kT_{c}=1.1\pm0.1$~keV for the soft photon source, a energy spectral
index of $\alpha=1.60\pm0.03$, and a geometric factor $f$,
parameterizing the fraction of the total soft photon flux illuminating
the BMC inflow atmosphere, of $f=0.32\pm0.02$. The observed
2-100~keV flux was $5.7\times 10^{-8}$ ergs/cm$^2$/s. We note
that the corresponding luminosity in the hard power-law component
is about $1.5\%$ of $L_{E}$ (with an assumption of the distance to 
the source, $3.2$ kpc and the mass of the central object,  7 solar masses),
which is consistent with our
assumption of negligible radiation pressure near the event
horizon.  Similar results; $kT_{c}=0.9\pm0.1$, $\alpha=1.68\pm0.03$  and
$f=0.72\pm0.02$, were obtained for GRS 1915+105.

From the inferred 2-200~keV luminosity for 
GRO~J1655-40, $\sim 5\%$ of $L_{E}$,
we derive a mass accretion rate (in Eddington units) of order 1,
bearing in mind the efficiency of gravitational to radiative 
energy conversion is of order 5\% or less, e.g. Shakura \& Sunyaev 1973
(hereafter SS73). This value of $\dot M$ is consistent with 
expected values within the BMC framework.
This suggests that the line-of-sight column density 
of the BMC atmosphere 
is of order $10^{24}$ cm$^{-2}$ (see, TMK97, Eq. 2).
However, because the best-fitted color temperature is about 1 keV, 
(and this is a lower bound on the BMC plasma temperature) we conclude
that the detected X-ray spectrum is not 
significantly modified by absorption (see also \S 2).
   
The temperatures we infer for these two sources are somewhat lower
than values previously reported [e.g. Zhang et al. (1997b)], however
this is to be expected. As noted, our procedure represents a fully
self-consistent model deconvolution, whereas most previous
approaches are phenomenological, i.e. power law plus black body or
multicolor disk. Mathematically, one expects the power law component
contribute significant soft-energy flux with the
net effect of skewing the thermal residual to higher apparent
temperatures. This will not occur with approach, as the
hard power law turns over towards low energies
(see Fig 3, TMK97). The inferred spectral indices also
agree extremely well with our model predictions. In TZ98,
calculations of the $\dot m - \alpha$ relationship were presented.
For the low temperature limit, an asymptotic lower limit of
$\alpha\simeq1.8$ was calculated; for the higher BMC plasma temperatures
(of order 10 keV)  this limit is significantly lower, $\sim1$, and for
mass accretion rates of  $\dot m\sim1$ (see below), $\alpha$
is precisely in the $1.5-1.8$ range we find (LT98, Titarchuk 1998).
Thus, we feel our observational test provides extremely
encouraging support of our methodology.

Using our inferred  color temperature $T_c$
and spectral index $\alpha$, along with the  measured flux
normalization, we can estimate the mass accretion rate, black hole
mass and source distance within the framework of standard accretion
disk theory (e.g. SS73). This additionally requires certain
assumptions  regarding a ''hardening factor"  -- the ratio of color
temperature to the effective plasma 
temperature (\cite{shimura95}, hereafter ShT95).
In fact, the spectral index (TMK97, TZ98, LT98) 
depends on the mass accretion rate and the plasma temperature;
the disk color temperature  is  $\propto (\dot m/m)^{1/4}$ (SS73),
and the normalization is $\propto \dot m m/d^2$ (where $d$ is 
the distance to the source). 

An ideal test case is GRO~J1655-40, since its mass and distance
are known to a high degree of accuracy (relative to other BHCs).
The distance we infer, $3.8\pm1.4$ kpc, is consistent (at the
$1-\sigma$ level) with previous
determinations (\cite{Hjellming95}). 
Also, the black hole mass we calculate
can be reconciled with determinations from dynamical
studies (\cite{orosz97}) provided we used the hardening factor 1.9
(ShT95). This assumed a mass accretion rate $\dot m=3$,
which was the value obtained from Monte Carlo simulations performed
to calculate the spectral index dependence on the mass accretion rate
and plasma temperature (LT98). Again,
this is an encouraging result suggesting that with further refinement,
one has a method of mass and distance determination
independent of the conventional quiescent spectroscopic and photometric
studies, which are not always plausible.

\section{DISCUSSION and CONCLUSION}
The successful application of our basic model to observational data and
the inferred physical parameters of those systems in comparison to
independent determinations is  encouraging.  We postulate that   (i){\it
The soft state detected in GRO J1655-40  and GRS 1915+105
represents a generic feature of accreting Galactic black holes. An
extragalactic analog may now be evident in the Narrow Line Seyfert 1
(NLS1) galaxy population.} (ii) {\it The BMC
spectra represent a characteristic signature of black hole horizons.
The disk flux, of order 5\% $L_{edd}$ tends to cool
the ambient environment and the generic hard power-law components 
is seen by the observer.}
(iii) {\it Because this spectral feature is formed very close to the
horizon, ($2-3R_s$), the variability timescales of high-energy line and
continuum radiation should be associated with the  crossing time scale
$t_{cross}\sim10^{-5}M/M_{\sun}$} s. (iv) {\it The  variability seen
in the soft component is not expected to be correlated with the hard
component. It is related to the illumination geometry of a small area of
the black hole horizon site, whereas  the soft radiation seen by the
observer directly emanates from a major fraction of the entire disk
which comprises a much larger area.}  (v) {\it QPOs emanating from
the inner edge of the accretion disk should lead to a pronounced hard-
X-ray variability signature, because the seed photons for the
converging flow upscattering come from the same inner disk region.}
 (vi) {\it The appearance  of  an  additional bump in the energy range 
10-20 keV can be explained in terms of downscattering (reflection) 
effects (ST80)  from the inner edge of the accretion disk.}

Our conclusions are consistent with various observations of Galactic
and extragalactic black hole systems. For example,  in NLS1s the X-ray
power-law is significantly steeper and its normalization is more
variable,  with time scale of order  $10^4$ s, than in broad-line
Seyfert 1 galaxies. This suggests the NLS1s may represent the
extragalactic analog of the high-soft state
(\cite{pounds95}, \cite{brandt97}, \cite{comastri98}). Several
groups have independently reached similar
conclusions (CT95, \cite{pounds95}).
We further note that the equivalent widths of Fe features detected  tend
to be large, in some cases $\sim500$ eV (\cite{comastri98}, hereafter
C98).

It is worth noting that the  detection of the strong hydrogen-like iron 
line is expected if the source of hard energy photons ($>7$ keV) 
is located inside the the converging inflow region(and alternatively, 
line radiation could form in the cooler, Compton cloud 
ambient to the converging
inflow region). It is easy to show that the ionization parameter
$\xi=L/(r^2n)\approx 10^5$ ergs~cm~$s^{-1}$ is a typical
value for the converging inflow and it is almost independent of the 
central object mass.
In this case, only the
hydrogen-like iron would be expected (Kallman and McCray 1982), 
which has been confirmed recently by BeppoSAX observations (C98).

Another supportive example is the black hole X-ray binary LMC X-3
which  appears to always be in the high state. Its hard-tail component
varies independently of the soft component
[e.g., \cite{ebisawa93}].

Recent RXTE observations of Cyg X-1 during a state
transition [\cite{cui97}] revealed a striking decrease in the
soft-to-hard photon lag times as the source passes from the hard to soft
state. This
is very strong empirical evidence that the soft seed  photons, which
comprise a of fraction the disk thermal component, and the hard-X-ray
power law emanate from a common compact region, again consistent
with our model.

The detection of 67 Hz QPOs  from GRS~1915+105 by RXTE was
recently reported by \cite{mrg97}.  It was  clearly demonstrated 
that this feature 
is  associated with  the high energy  component visible 
in the PCA. This
can be explained in terms of a QPO in the inner edge of the disk
or by $g-$mode disk oscillations occurring
within the characteristic radius of 4 $r_s$ [\cite{tlm98}].
This should lead to variations in the hard spectral component
since significant changes in the illumination geometry  of the
converging inflow site, can
occur (TMK97, TZ98).  Similar intrepretation can be
applied to the  300 Hz QPOs detected by RXTE in
GROJ1655--40 [\cite{rem97}].

 In conclusion,  we wish to emphasize once again that the observations
presented here, along with some observational trends presented by
others in the literature, and the relativistic theory prompt us to claim
that {\it we have identified a generic spectral signature black hole
accretion.}

\vspace{0.2in}

\centerline{\bf{ ACKNOWLEDGMENTS}}
We wish to acknowledge the anonymous referee for who made a number
of useful comments on the initial draft, as well as 
Jean Swank and Menas Kafatos for discussion and useful suggestions.
Portions of this work were supported by the Rossi X--Ray Timing 
Explorer and Compton Gamma Ray Observatory Guest Observer Programs.
L.T. also would like to acknowledge support from NASA grant NAG5-4965.

\newpage


\vspace{10pt}
{\bf Figure 1.} RXTE/PCA and BATSE/LAD
spectrum for
(a) Broad-band high-energy spectrum of GRO~J1655-40 obtained on 20
June, 1996. The data are fitted to the bulk--motion accretion model. The
resulting model parameters are as stated in the text.  (b) Same
instrumentation and methodology applied to November 1996 active state
in GRS~1915+105
%
%


\newpage

\begin{thebibliography}{}

\bibitem[Bandyophadyay et al. (1998)]{bandy98}
Bandyophadyay, R., Martini, P., Gerard, E.  Charles, P.A., Wagner, R.M.,
Shrader, C., Shahbaz,  T., \& Mirabel, I.F. 1998, MNRAS, in press

\bibitem[Brandt,  Mathur \&  Elvis (1997)]{brandt97}
 Brandt, W.N., Mathur, S. \&  Elvis, M. 1997, MNRAS, 285, 25

\bibitem[Chakrabarti \& Titarchuk (1995)]{ct95}
Chakrabarti S.K. \& Titarchuk, L. G. 1995, ApJ,  455, 623 (CT95)

\bibitem[Chaty et al. (1996)]{chaty96}
Chaty, S., Mirabel, I. F., Duc, P. A., Wink,  J. E., Rodriguez, L. F.,
1996, A\&A, 310, 825

\bibitem[Chen et al. (1997)]{csl97}
Chen, W., Shrader, C.R., \& Livio, M.,  1997, ApJ, 491, 312

\bibitem[Comastri et al. (1998)]{comastri98} 
Comastri, A., Fiore, F.,  et al. 1998, A\&A, in press (C98)

\bibitem[Cui et al. (1997)]{cui97}
 Cui, W., Zhang, S.N., Focke, W., \& Swank, J.H., 1997, ApJ, 484, 383

\bibitem[Ebisawa et al. (1993)]{ebisawa93}
Ebisawa, K., et al, 1993, ApJ, 403, 684

\bibitem[Ebisawa et al. (1994)]{ebisawa94}
Ebisawa, K., et al, 1994, PASJ, 46, 375

\bibitem[Esin,  McClintock \&  Narayan,  (1997)]{esin97}
 Esin, A.A., McClintock, J.E., \&  Narayan, R., 1997, ApJ, 489, 865

\bibitem[Fabian et al. (1995)]{fabian95}   Fabian, A. C., Nandra, K.,
Reynolds, C. S., Brandt,  W. N., Otani, C., Tanaka, Y., Inoue, H., Iwasawa, K.
1995, MNRAS, 277, L1

\bibitem[Hjellming,\& Rupen (1995)]{Hjellming95} Hjellming, R.M., \&
 Rupen, M.P., 1995, {it Nature}, 375, 464.

\bibitem[Hynes et al. (1998)]{hynes98}  Hynes, R.I., Haswell, C. A.,
  Chen, W., \&  Shrader, C. R., 1998, MNRAS, in press

\bibitem[Inoue et al. (1994)]{inoue94}
 Inoue, H. et al., 1994, IAUC 6063

\bibitem [Kallman \& McCray (1982)]{kal82}
Kallman, T.R. \& McCray, R. 1982, ApJS, 50, 263

\bibitem[Laurent \& Titarchuk  (1998) ]{lt98}
Laurent, Ph., \& Titarchuk, L.G., 1998 in preparation (LT98)

\bibitem[Liang \& Narayan (1997)]{liang97}
Liang, E., \& Narayan, R., 1997, in
{\it Proceedings of the Fourth Compton Symposium} (eds Dermer,
C.D. Strickman, M.S., \& Kurfess, J.D.), AIP CP-410

\bibitem[Morgan, Remillard \&  Greiner (1997)]{mrg97}
 Morgan, E.H., Remillard, R.A., \&  Greiner, J., 1997, ApJ, 482, 993

\bibitem[Narayan, Garcia \& McClintock,  (1997)]{narayan96}
 Narayan, R., Garcia, M.R., \& McClintock, J.E.,  1997, ApJ, 478, L79

\bibitem [Orosz \& Bailyn (1997)]{orosz97}
Orosz, J., \& Bailyn, C.D.  1997, ApJ, 477, 876

\bibitem[Pounds, Done \& Osborne  (1995)]{pounds95}
 Pounds, K.A., Done, C., \& Osborne, J.,  1995, MNRAS, 277, L5

\bibitem[Remillard et al. (1997)]{rem97}
Remillard, R.A., Morgan, E.H., McClintock, J.E., Bailyn, C.D.
 Orosz, J.A., \& Greiner, J., 1997, (astro-ph/9705064)

\bibitem[Shakura \& Sunyaev (1973)]{ss73}
Shakura, N.I., \& Sunyaev, R.A., 1973,  A\&A, 24, 337 (SS73)

\bibitem [Shimura \& Takahara (1995)]{shimura95}
 Shimura, T., \& Takahara, F., 1995, ApJ, 445, 780 (ShT95)

\bibitem[Sunyaev \& Titarchuk (1980)]{st80}
Sunyaev, R.A., \& Titarchuk, L.G. 1980, A\&A,  86, 121 (ST80)

\bibitem[Tanaka et al. 1995]{tanaka95}
 Tanaka, Y., Nandra, K., Fabian, A. C., Inoue, H.,
 Otani, C., Dotani, T., Hayashida, K., Iwasawa, K., Kii, T., Kunieda,
H.,  Makino, F., Matsuoka, M., 1995,  Nature 375, No.6533, 659

\bibitem[Titarchuk \& Zannias (1998)]{tz98}
Titarchuk, L., \& Zannias. T., 1998, ApJ, 493, 863 (TZ98)

\bibitem[Titarchuk, Mastichiadis \&  Kylafis, (1997)]{tmk97}
Titarchuk, L. G., Mastichiadis, A.,  Kylafis, N. D. 1997, ApJ, 487, 834 (TMK97)

\bibitem [Titarchuk, Lapidus \& Muslimov (1998)]{tlm98}
 Titarchuk, L.G., Lapidus, I.I. \& Muslimov, A., 1998,
 ApJ, 499, (astro/ph-9712348)

\bibitem[Titarchuk  (1998) ]{tit98}
 Titarchuk, L.G., 1998 in preparation

\bibitem[Zhang et al. (1997a)]{zhang97a}  Zhang, S.N., Mirabel, I.F.,
Harmon, B.A.,  Kroeger, R.A., Rodriguez, L.F., Hjellming, R.M., \& Rupen, M.P.,
1997a  in {\it Proceedings of the Fourth Compton Symposium} (eds Dermer,
C.D. Strickman, M.S., \& Kurfess, J.D.), AIP CP-410

\bibitem[Zhang et al. (1997b)]{zhang97b}
Zhang, S.N., et al, 1997b, ApJ, 479, 381


\end{thebibliography}
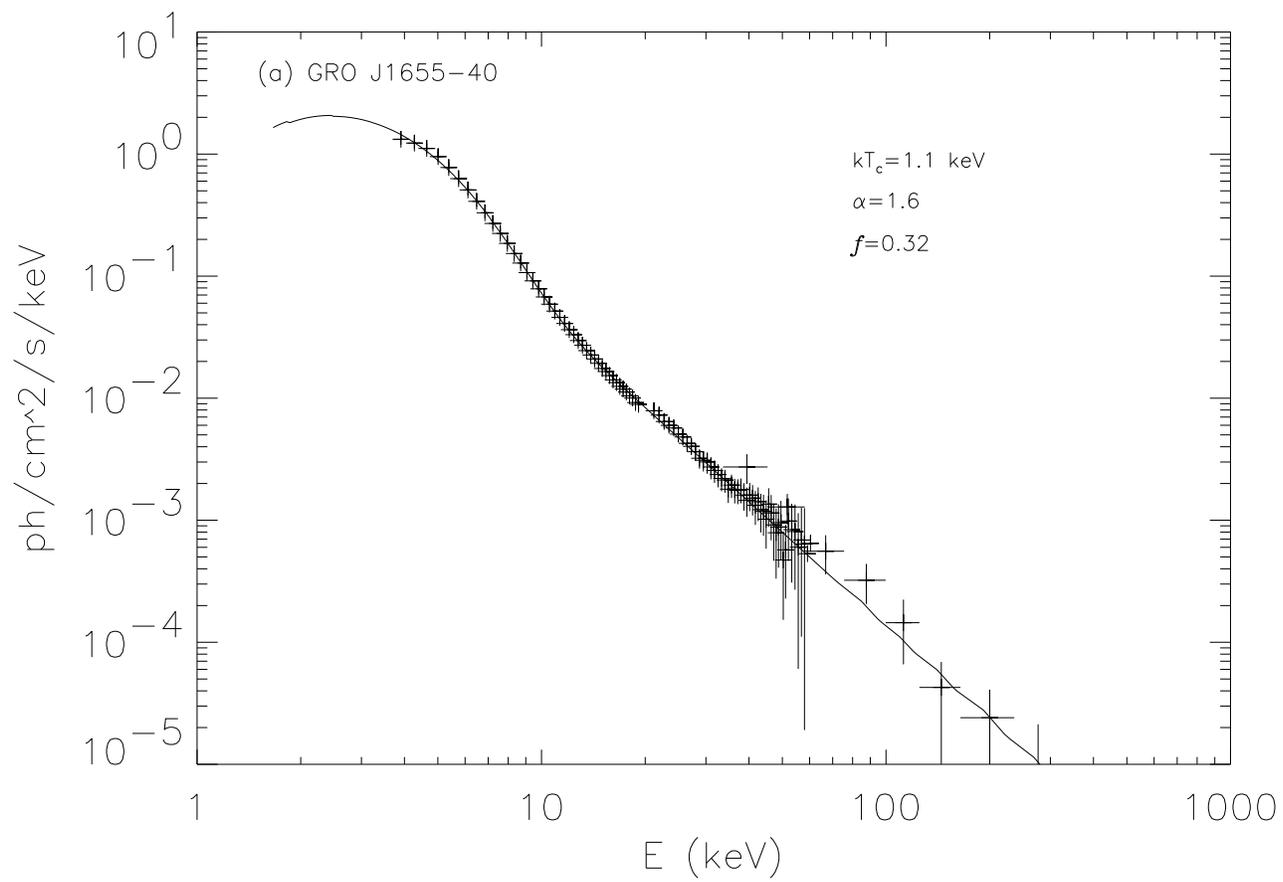
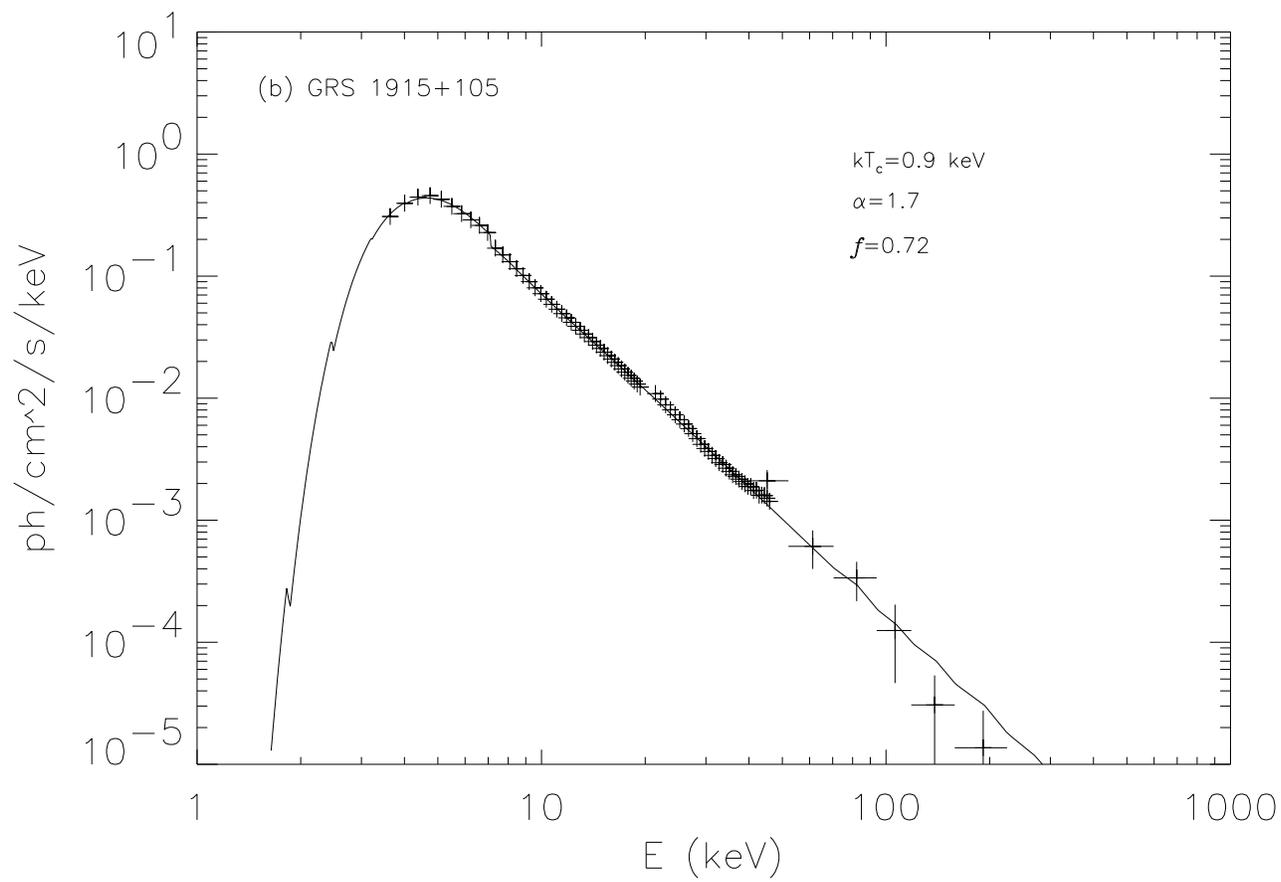
\end{document}